\documentstyle[eqsecnum,floats,preprint,epsf,aps]{revtex}
\tighten

\begin{document}

\renewcommand{\arraystretch}{1.5}
\newcommand{\be}{\begin{equation}}
\newcommand{\ee}{\end{equation}}
\newcommand{\bea}{\begin{eqnarray}}
\newcommand{\eea}{\end{eqnarray}}
\def\Tr{\mathop{\rm Tr}\nolimits}
\def\mapright#1{\smash{\mathop{\longrightarrow}\limits^{#1}}}
\def\mapdown#1{\big\downarrow \rlap{$\vcenter
  {\hbox{$\scriptstyle#1$}}$}}
\def\su#1{{\rm SU}(#1)}
\def\so#1{{\rm SO}(#1)}
\def\sp#1{{\rm Sp}(#1)}
\def\u#1{{\rm U}(#1)}
\def\o#1{{\rm O}(#1)}
\def\p#1{{\pi_{#1}}} 
\def\Z{{\bf Z}}
\def\R{{\bf R}}
\def\M{{\cal M}}
\def\L{{\cal L}}
\def\N{{\cal N}}
\newcommand{\PSbox}[3]{\mbox{\rule{0in}{#3}\includegraphics{#1}\hspace{#2}}}

\title{Domain Wall Junctions are $1/4$-BPS States}

\author{Sean M. Carroll$^{1}$\footnote{Address after 1 July 1999:
Enrico Fermi Institute, University of Chicago, 5640 S.~Ellis Avenue, 
Chicago, IL~60637; {\tt carroll@theory.uchicago.edu}}, 
Simeon Hellerman$^{2}$, and Mark Trodden$^{3}$}

\address{~\\$^1$Institute for Theoretical Physics \\
University of California \\
Santa Barbara, California 93106, USA.\\
{\tt carroll@itp.ucsb.edu}}

\address{~\\$^2$Department of Physics \\
University of California \\
Santa Barbara, California 93106, USA.\\
{\tt sheller@physics.ucsb.edu}}

\address{~\\$^3$Department of Physics \\
Case Western Reserve University \\
10900 Euclid Avenue \\
Cleveland, OH 44106-7079, USA.\\
{\tt trodden@theory1.physics.cwru.edu}}

\maketitle

\begin{abstract}
We study $N=1$ SUSY theories in four dimensions with multiple discrete
vacua, which admit solitonic solutions describing segments of domain
walls meeting at one-dimensional junctions. We show that there exist
solutions preserving one quarter of the underlying
supersymmetry -- a single Hermitian supercharge.  We derive a BPS
bound for the masses of these solutions and construct a solution
explicitly in a special case.  The relevance to the confining phase of
$N=1$ SUSY Yang-Mills and the M-theory/SYM relationship is discussed.

\end{abstract}

\setcounter{page}{0}
\thispagestyle{empty}

\vfill


\noindent CWRU-P21-99\hfill hep-th/9905217

\vfill
\eject

\baselineskip 20pt plus 2pt minus 2pt

\section{Introduction}
\label{introduction}

Duality has played a fundamental role in recent progress in
understanding quantum field theories at strong coupling.  One of the
most indispensable tools in formulating and providing evidence for
duality conjectures for supersymmetric theories has been the existence
of states preserving some but not all of the underlying supersymmetry
of the Hamiltonian.  These states, referred to as BPS states, are
useful because they lie in shortened multiplets of the supersymmetry
algebra, and therefore they cannot disappear or appear as parameters
of the theory, such as the coupling constant, are varied continuously.
The spectrum of BPS states is thus one of the few characteristics of a
quantum field theory that can be predicted easily at strong coupling.

A celebrated application of this tool has been to gauge theories in
four dimensions with extended ($N=2$, 4) supersymmetry.  These
theories feature particle-like solitons --- magnetic monopoles ---
that preserve half the supersymmetry of the Hamiltonian, lie in short
multiplets, and therefore can be followed to strong coupling, where
they become the fundamental excitations of a dual, weakly coupled
theory.

In $N=1$ theories in four dimensions, the situation is different.  The
SUSY algebra forbids a rotationally invariant central charge and thus
massive zero-dimensional objects, such as magnetic monopoles, cannot lie in
short representations of the algebra.  However, it has been noted
that the $N=1$ SUSY algebra in 4D can admit a central
extension if the central charges transform nontrivially under the 
rotation group \cite{a&t,dvalshif}.  
Consequently, in such theories there can exist states which
are extended objects preserving half the supersymmetry --- two
Hermitian supercharges.  These objects are domain walls
separating two of a set of disconnected vacua.

In this paper, we show that, if an $N=1$ theory has three 
or more mutually disconnected vacua,
there exist states preserving a \it quarter \rm of the underlying
supersymmetry --- a single Hermitian supercharge.  These states are
junctions of BPS domain walls.  Networks of intersecting walls
have been studied in a cosmological context \cite{rps}, and 
theories supporting topological defects ending on other defects 
of various dimensions (including models of walls ending on walls)
have also been constructed \cite{ct}.  In addition, reference \cite{t&v}
contains a general discussion of domain wall intersections in $N=1$ theories.

The organization of the paper is as follows.  In section
\ref{defects}, we give a brief review of BPS domain walls in $N=1$
theories.  In section \ref{bps}, we find the BPS equations satisfied
by a soliton preserving a quarter of the underlying supersymmetry.  In
sections \ref{z3} and \ref{ldl} we use the BPS equations to derive some
constraints on the kinematics of these solutions, and in section
\ref{z4} we use what we've learned to construct solutions.
In section \ref{sym}
we then apply these considerations to the particular case of $N=1$
supersymmetric Yang-Mills theory, before concluding and discussing
open questions in section \ref{discussion}.

As this work was being completed, we became aware of a related paper
by Gibbons and Townsend \cite{gt}, in which they also argue on general
grounds for the existence of BPS
wall junctions preserving $1/4$ of the $N=1$ supersymmetry.

\section{Supersymmetric Domain Walls}
\label{defects}

To understand the origin of central charges in $N=1$ theories with only chiral
superfields, consider the theory with superpotential 

\be
  W = \Lambda^2{\Phi} - \Lambda^{2-n}{{\Phi^{n+1}}\over{n+1}} \ ,
  \label{super}
\ee
where $n$ is an integer $\geq 2$.
The anticommutator $\{Q_\alpha, Q_\beta\}$ of two supercharges 
in this theory doesn't automatically vanish, 
but, for a static configuration,  is proportional to the
integral of the total derivative

\bea
\epsilon_{\alpha\gamma} 
(\sigma_0 \bar{\sigma}^a)_\beta^\gamma \partial_a W(\phi)^* \nonumber \ ,
\eea
where $a$ runs over spacelike indices and $\phi$ is the scalar 
component of the superfield $\Phi$.
The supersymmetry algebra therefore closes on a
non-scalar central charge $Z^a$, which is proportional to the change in
the value of the superpotential between spatial infinities in different
directions.  This extension of the SUSY algebra
\cite{a&t,dvalshif} is relevant to the confining phase of pure supersymmetric
gluodynamics in 4 dimensions, where it has been shown \cite{witcon} that,
for the gauge group $SU(n)$,
the theory has $n$ distinct vacua corresponding to distinct values of the 
superpotential.  It was shown in \cite{dvalshif} that this
central charge allowed the existence of $1/2$-supersymmetric domain walls,
where the order parameter which changes across the wall
is essentially the expectation value of the gluino condensate.

More concretely, consider the full SUSY algebra

\bea
  \{Q_\alpha, \bar{Q}_{\dot{\beta}}\} & = & 2 \sigma^\mu_{\alpha\dot{\beta}}
   P_\mu
  \label{wallanom} \\
  \{Q_\alpha, Q_\beta\} & = & - 2 Z^a \epsilon_{\alpha\gamma}
  (\sigma_0 \bar{\sigma}^a)_\beta^\gamma  \ , 
\eea
where $P_\mu$ is the energy-momentum vector of the system.  We see that a 
state with energy $H = (Z^a Z^{*a})^{1/2}$
preserves the supercharges 

\be 
  Q_\alpha - \epsilon_{\alpha\gamma}\bar{\sigma}^{a\dot{\beta}\gamma} 
  {{Z^a}\over{ (Z^b Z^{*b})^{1/2} }} \bar{Q}_{\dot{\beta}}
\ee
and their Hermitian conjugates.
Notice that, although this appears to give four Hermitian
supercharges, there are really only two linearly independent Hermitian
supercharges unbroken.

\section{BPS bounds for junctions}
\label{bps}

We now turn to the general BPS properties of the wall junctions we have just
described, and those we will introduce later.
The $N=1$ chiral theory of the previous section has supercurrent  
\be
  j_{\mu\alpha} = i\sqrt{2}(\sigma_\mu\bar{\sigma}^\nu)_\alpha^\beta 
\partial_\nu \phi^*
  \cdot \psi_\beta 
+ i \sqrt{2} W^\prime(\phi)^* \sigma_{\mu\alpha\dot{\gamma}}
  \bar{\psi}^{\dot{\gamma}} \ ,
\ee
with associated supercharge 
\be
Q_\alpha = i \sqrt{2} \int d^3 x \left[ 
(\sigma^0 \bar{\sigma}^\nu)_\alpha^\beta \partial_\nu \phi^*(\vec{x})\cdot
\psi_\beta(\vec{x})+ W^\prime(\phi(\vec{x}))^*
 \sigma^0_{\alpha\dot{\gamma}}\bar{\psi}
^{\dot{\gamma}}(\vec{x})\right] \ .
\ee
A careful calculation of the anticommutator $\{Q_\alpha, \bar{Q}_{\dot
{\beta}}\}$ recovers not only the usual momentum term, but also a total
derivative given by $(\sigma^a_{\alpha\dot{\beta}} \cdot Y^a)$, where
\be
Y^a \propto \epsilon^{abc}\int d^3 x \left[\partial_b\phi( \vec{x} ) 
\partial_c\phi^*( \vec{x} ) + \rm{h.c.} \right]\ .
\label{juncanom}
\ee
A similar central term arises in theories which admit supersymmetric
string solutions, including many supergravity theories \cite{extend}, and
$N=1$ SUSY gauge theories with abelian gauge group factors and non-vanishing
F-I parameters \cite{cosmic}.  Static string (or multi-string) solutions
have a tension determined by the value of the central charge $Y^a$, and
preserve half the supersymmetry of the Hamiltonian.

If analogous partially supersymmetric states exist in the theory we are 
considering,
then they must look very different, since clearly the Lagrangian
cannot admit string solutions with finite tension. We would like to know
whether there are any
BPS states in chiral $N=1$ theories with a nonzero value of the central term
$Y^a$.

As we shall see in the rest of this paper, the answer is yes.  We find
that in theories with only chiral superfields, the central term $Y^a$
admits an interpretation not as string charge but rather as {\it
junction charge}.  In any 4D field theory with more than two
disconnected vacua, the domain walls may meet in one-dimensional
junctions.  If the theory is supersymmetric, we demonstrate that
junctions of the 1/2-supersymmetric domain walls may have stable
junction solutions preserving 1/4 of the supersymmetry of the
Hamiltonian.

{}From this point onwards, we will assume the junction state to be
static and translationally invariant in the direction $x_3$.  For such a
configuration, the magnitude of the central term $Y^a$ plays the role
of an additional contribution to the mass of a junction, above and
beyond that contributed by the half-walls themselves.  That is, not
only do the ``spokes'' of the junction have a tension associated with
them, but the ``hub'' has its own non-vanishing contribution to the
total energy, in contrast to the string junctions of
\cite{stringjunc}.

As in the more familiar examples of central charges in SUSY algebras,
the central term here arises at the semiclassical level as a topological term
entering a classical BPS bound.  To see this, note that the Hamiltonian for 
static configurations,

\be
H = \int d^3x \, \left[
(\partial_{x_1} \phi ) (\partial_{x_1} \phi^*) + 
(\partial_{x_2} \phi ) (\partial_{x_2} \phi^*) + 
(\partial_{x_3} \phi ) (\partial_{x_3} \phi^*)  + W^\prime(\phi)^* 
W^\prime(\phi)\right] \ ,
\ee
can be rewritten, for any phase $\Omega$, as the sum of positive definite 
terms and a total derivative term:

\bea
H & = \int d^3x \, & \left[(\partial_{x_3} \phi ) (\partial_{x_3} \phi^*)
+ (\partial_{x_1} \phi  - i \partial_{x_2}\phi - \Omega W^\prime(\phi)^*)
  (\partial_{x_1} \phi^*  + i \partial_{x_2}\phi^* - \Omega^* W^\prime(\phi))
\right.
\nonumber \\
& & + \left.(\partial_{x_1} - i \partial_{x_2} )( \Omega^* W)  
+ (\partial_{x_1} + i \partial_{x_2})(\Omega W^*) 
+ i \partial_{x_1} \phi^* \partial_{x^2} \phi - i \partial_{x_1} \phi
\partial_{x_2} \phi^*\right] \nonumber \\
& = \int d^3x \, & \left[(\partial_{x_3} \phi ) (\partial_{x_3} \phi^*)
+ 4 (\partial_z \phi - {1\over 2} \Omega W^\prime(\phi)^*)
  (\partial_{\bar{z}}\phi^* - {1\over 2} \Omega^* W^\prime(\phi))
\right. \nonumber \\
& & + 2 \left.\partial_z ( \Omega^* W)  
+ 2 \partial_{\bar{z}} (\Omega W^*) 
+ \partial_z (\phi^* \partial_{\bar{z}} \phi - \phi \partial_{\bar{z}} 
\phi^*) 
+ \partial_{\bar{z}} (\phi \partial_z \phi^* - \phi^* \partial_z \phi)\right]
\label{clasbps}
\eea

Since the mass in any given region is equal to a positive definite
term plus a surface term, this imposes a classical BPS lower bound on
the mass in a region in terms of the values of the fields on its
boundary.
If the positive definite terms are set to zero (which will turn out
precisely to impose
the BPS equations for a static configuration) then the total mass of
the state becomes a surface term.  Upon doing the integral we find that
the first pair of surface terms (involving the superpotential) gives the
contribution to the mass corresponding to the central charge $Z^a$, and the
second pair gives the contribution corresponding to the central charge
$Y^a$.

For later convenience, note that this calculation is easily generalized to
the case of a nontrivial K\"ahler metric $K_{\phi \phi^*}$.  In that case, 
the relevant surface terms in the energy are
\be
 2 \partial_z ( \Omega^* W)  
+ 2 \partial_{\bar{z}} (\Omega W^*) 
+ \partial_z (K_{,\phi} \partial_{\bar{z}} \phi - K_{,\phi^*}
 \partial_{\bar{z}} \phi^*) 
+ \partial_{\bar{z}} (K_{,\phi^*} \partial_z \phi^* - K_{,\phi}
 \partial_z \phi)\ .
\ee
Therefore, the mass contributed by the junction charge $Y^a$ is equal
to a line integral of the pullback of the one-form $i K_{,\phi} d\phi
- i K_{,\phi^*} d\phi^* $, or, equivalently:
\\
\\
{\it The junction mass is proportional to the area in field space spanned by
the fields of the solution, as measured by the K\"ahler metric.}
\\

Although, unlike
the masses of the domain walls themselves, the mass of the junction is not
protected by supersymmetry from perturbative quantum corrections,
this is still a compact and useful result, and we shall use it later to 
derive an interesting quantitative prediction about the behavior of certain 
states in M-theory.

\section{Domain wall junctions: the $Z_3$ case}
\label{z3}

\begin{figure}
  \centerline{\epsfbox{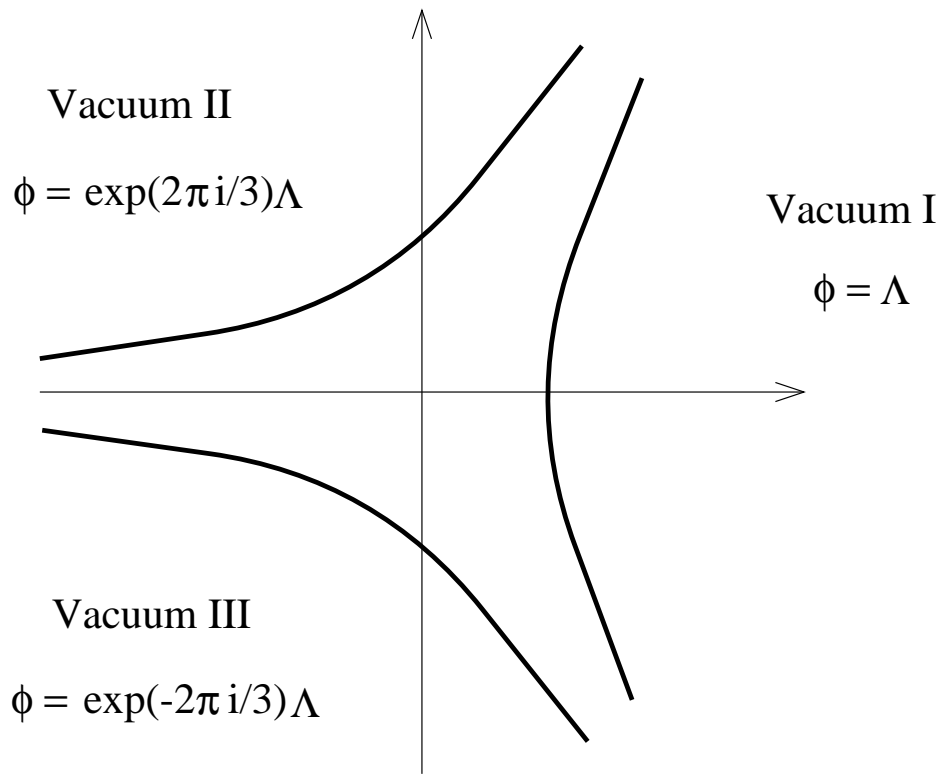}}
  \vspace{0.3in}
  \caption{\sf A field configuration that interpolates among the three 
   supersymmetric vacua.  The energy density is concentrated inside the
   contour.}
  \label{stwall1}
\end{figure}

In order understand the subtleties involved in constructing our $1/4$-BPS 
states, let us specialize to the simplest nontrivial example, the case $n=3$.
The superpotential (\ref{super}) becomes
\be
  W = \Lambda^2 \Phi - {{\Phi^4}\over{4\Lambda}} \ ,
\label{supot3}
\ee  
so that the theory has three supersymmetric vacua $I$, $II$, and $III$, 
in which the scalar component takes on vacuum expectation values
\be
  \phi_I = \Lambda,\quad \phi_{II} = e^{{{2\pi i}/ 3}} \Lambda\ ,
  \quad \phi_{III} = e^{-{{2\pi i}/ 3}} \Lambda\ .
\ee
We wish to consider an initial field configuration in which
$\phi$ tends to each of these values in three different directions,
as in figure \ref{stwall1}.
If we allow the field to radiate away energy, it will settle down to a
stable configuration whose topology is that of a junction of three half-walls
meeting at $120^\circ$ angles.

We now analyze how much supersymmetry may be preserved by such a 
configuration.  First, consider the supercharges
left unbroken by each individual half-wall.  The changes
$\vec\Delta W \equiv \int d^3x\, \vec\nabla W$ in the 
superpotential across the walls between vacua $I$ and
$II$, between vacua $II$ and $III$, and between vacua $III$ and $I$, are
\bea
\vec{\Delta} W_{I, II} & = & \left(- {{\sqrt{3}}\over 2} , {1\over 2} , 
0\right)\cdot
\left(\exp{{{2\pi i}\over 3}} - 1\right)\cdot{{3\Lambda^3}\over 4} \\
\vec{\Delta} W_{II, III} & = & \left(0 , -1 , 0\right)\cdot
\left(\exp{{{4\pi i}\over 3}} - \exp{{{2\pi i}\over 3}} \right)
\cdot{{3\Lambda^3}\over 4}
\\
\vec{\Delta} W_{III, I} & = & \left(- {{\sqrt{3}}\over 2} , {1\over 2} , 
0\right)\cdot
\left(1- \exp{{{4\pi i}\over 3}} \right)\cdot{{3\Lambda^3}\over 4} \ ,
\eea
respectively.  Therefore, the supercharges preserved by each half-wall 
alone are
\bea
Q_{I, II}^{(1)} & = & Q_\uparrow - Q_\uparrow^\dagger \\
Q_{I, II}^{(2)} & = & Q_\downarrow + Q_\downarrow^\dagger \\
Q_{II,III}^{(1)} & = & Q_\uparrow - Q_\uparrow^\dagger \\
Q_{II,III}^{(2)} & = & Q_\downarrow + \exp\left\{ - {{2\pi i}\over 3} \right\} 
Q_\downarrow^\dagger \\
Q_{III,I}^{(1)} & = & Q_\uparrow - Q_\uparrow^\dagger \\
Q_{II,III}^{(2)} & = & Q_\downarrow + \exp\left\{ + {{2\pi i}\over 3} \right\} 
Q_\downarrow^\dagger \ .
\eea
Here we have written the conjugates of $Q$ in terms of $Q^\dagger$ instead
of $\bar{Q}$ as we are dealing with static states, and wish to emphasize
Hermiticity rather than Lorentz covariance.  The subscripts $\uparrow,
\downarrow$ represent the supercharges with spin $\pm {1\over 2}$ in the
$z$-direction, respectively. Thus the entire configuration can preserve at most
the Hermitian supercharge
\be
Q_{junc} = i (  Q_\uparrow - Q_\uparrow^\dagger )
\ee

The equations satisfied by a semiclassical solitonic junction state
$|J\rangle$ preserving this supercharge can be found by taking the
expectation value of the SUSY variations of the fermions:
\bea
0 & = & - i \langle J|\{Q_{junc} , \psi_\downarrow \} |J\rangle \nonumber \\ 
& = & \langle J | (\partial_x - i \partial_y) \phi - W^\prime(\phi^*)|J\rangle
\nonumber \\
& \simeq & (\partial_x - i \partial_y) \langle J|\phi|J\rangle - W^\prime
(\langle J|\phi^*|J\rangle) \ , \\
& & \nonumber \\
0 & = & \langle J|\{Q_{junc} , \psi_\uparrow \} |J\rangle  \nonumber \\
& = & \langle J| (\dot{\phi} + \partial_z \phi )|J\rangle \ ,
\eea
in the $\hbar \to 0$ limit.  Notice that these are exactly the equations
obtained by assuming saturation of the classical BPS bound
(\ref{clasbps}).  The existence of $1/4$-BPS domain wall junctions
depends on the existence of a solution to these equations. However, in 
general it is quite difficult to solve the equations directly.

In order to address this, we pursue two different strategies.  First,
in section \ref{ldl} we present a picture of wall junctions and
junction
networks viewed on scales large compared to the thickness of the walls,
with emphasis on the $Z_3$ case. 
We derive a new form of the BPS equations appropriate to this regime,
and use it to show that the BPS equations translate into a condition on the
kinematics of the junctions.

In section \ref{z4} we take a different tack, and
outline an explicit construction for junction and network solutions to the
BPS equations.  We apply this approach to a particular intersecting domain wall
solution, in which two BPS
walls intersect, rather than one terminating on the other.  While we do
not resolve the existence question for domain wall junctions in general, the
construction demonstrates that 1/4-BPS solutions do indeed exist in many
cases.

\section{Long distance limit and integral BPS equations}
\label{ldl}

We will begin with the long-distance limit. Our strategy is to derive an
integral, global form of the BPS equations and use this to examine domain
wall junctions on scales much larger than the thickness of the walls.  We will
see that in this long-distance regime, the BPS condition is indeed satisfied
for various configurations. 

We consider field configurations which
tend to some vacuum everywhere at spatial infinity, except perhaps along
codimension-one defects separating different vacua.  Such configurations have
approximately step-function behavior across domain walls, when viewed
on scales much larger than any length scale appearing in the Lagrangian.

\begin{figure}
  \centerline{\epsfbox{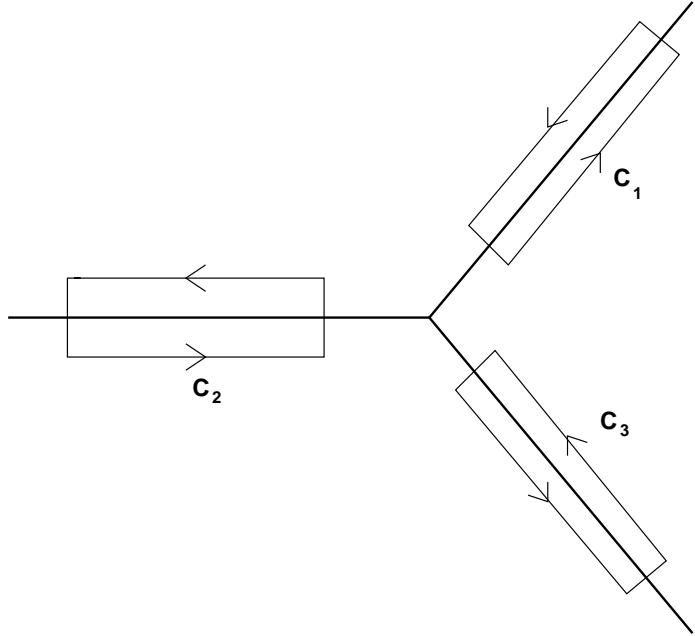}}
  \vspace{0.3in}
  \caption{\sf A long-distance view of a three-wall junction, and three
   elongated rectangular contours in the $z-\bar{z}$ plane.}
  \label{longdist1}
\end{figure}

How do we check to see whether or not the BPS equations are 
satisfied for a given configuration in this limit?  Since we want to work in
an approximation in which the fields vary
discontinuously, the differential form of the BPS equations is clearly
unsatisfactory.  However we can write down an integral form of the
equations, analogous to the integral form of the first-order Maxwell
equations.

We begin by noting that the BPS equations
\be
2 K_{\phi \phi^*} \partial_z \phi = \Omega W^\prime(\phi)^*
\ee
and
\be
2 K_{\phi \phi^*} \partial_{\bar{z}} \phi^* = \Omega^* W^\prime(\phi)
\ee
imply the relations
\be
\partial_z W(\phi)  = W^\prime (\phi) \partial_z \phi = {\Omega\over 2}
K^{\phi\phi^*}  W^\prime (\phi) W^\prime(\phi^*) = {\Omega\over 2}
V(\phi,\phi^*) 
\ee
and
\be
\partial_{\bar{z}} W(\phi)^* = {{\Omega^*}\over 2}
V(\phi,\phi^*) \ ,
\ee
where $V$ denotes the potential energy.
We then integrate these equations over any large region $R$ in the
$z$-$\bar{z}$ plane; in particular, integrate over an elongated
rectangle containing a segment of one of the walls, whose long side
has length $L >> \Lambda^{-1}$.

Using Stokes's theorem,
\be
\int_R dz \wedge d\bar{z} (\partial_z v_{\bar{z}} - \partial_{\bar{z}} v_z)
= \oint_{C\equiv \partial R} (dz v_z + d\bar{z} v_{\bar{z}})\ ,
\ee
and setting
\bea
v_{\bar{z}} & \equiv & - i \Omega^* W \nonumber \\
v_z & = & i \Omega W^* \ , 
\eea
we obtain
\be
- i \int_R dz \wedge d\bar{z} V = i \oint_C (\Omega W^* dz - \Omega^* W d
\bar{z})\ .
\ee

It is straightforward to check that for an isolated BPS wall, the kinetic
and potential energies of the solution are equal, and so this must hold
as well for the junction solution to order $L^1$.
Since $dz \wedge d\bar{z} = -2i dx\wedge dy$, the equation above
then says that the total mass
enclosed by the contour $C$, for a 1/4-supersymmetric
junction state, is given to order $L^1$ by the contour integral

\bea
 M_{\rm enclosed} = - i \oint_C (\Omega W^* dz - \Omega^* W d
\bar{z})  \ .
\eea

This reformulation of the BPS equations immediately yields a useful set of
restrictions on the statics of wall junctions.  Assuming the walls themselves
to saturate the BPS bound
\be
T = 2 |\Delta W |,
\ee
taking $L\to \infty$ and matching terms of order $L^ 1$,
we find that each wall must be oriented such that $\Omega \Delta W \omega$
is real and positive, where $\omega$ is the phase characterizing the
orientation of the wall in the complex plane.

\begin{figure}
  \centerline{\epsfbox{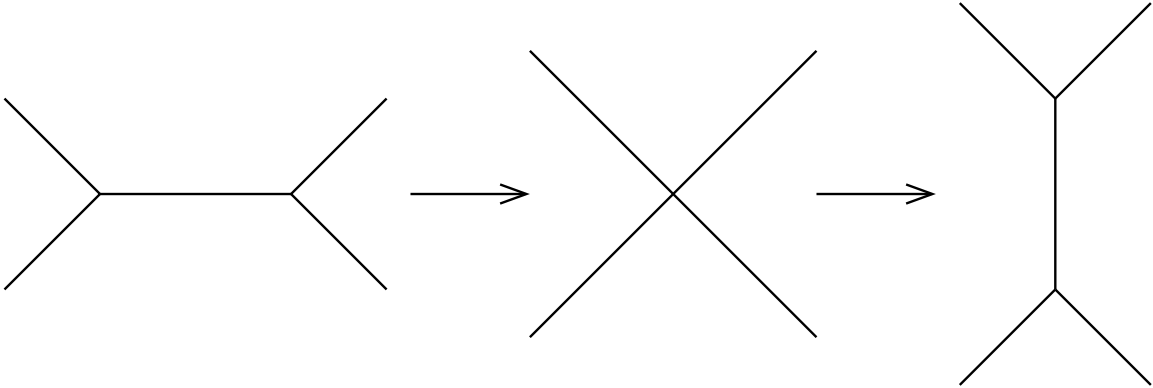}}
  \vspace{0.3in}
  \caption{\sf In the $Z_4$ theory,
   the long-distance limit suggests the existence of a
   one-dimensional moduli space of BPS wall junction networks with four
   external walls.  The two branches of moduli space, which
   meet at a $Z_4$-symmetric branch point, resemble t- and
   s-channel Feynman diagrams, respectively.}
  \label{moduli}
\end{figure}

Thus, since $\Delta W_{I,II}$, $\Delta W_{II,III}$, and $\Delta
W_{III,I}$ differ from each other in phase by $120^0$, this means that
the orientations of the walls in the $z$-$\bar{z}$ plane must also
differ by the same amount.  Therefore, the integral forms of the BPS
equations confirm reasonable physical expectations for the angles at
which the walls must meet, based on the balance of forces on the
junction.  Although the $Z_3$ symmetry of this particular case makes
the statics particularly simple, it is straightforward to check that
the integral form of the BPS equations yields the same consistent
picture for an arbitrary superpotential with multiple vacua: the
relative angles of walls at a junction must be arranged so as to
cancel the total force on the junction point.
Therefore, for theories with four or more
disconnected supersymmetric vacua, we argue that unless there
may to exist full moduli spaces of domain wall networks,
as in figure \ref{moduli}.  We caution, however, that the moduli space
may not exist beyond this limit; the low-frequency dynamics on this space
are governed by a $1+1$-dimensional field theory with only a single
supercharge, which allows for the existence of potentials.  The question
of the existence or nonexistence of such potentials lies beyond the
scope of this paper.

It remains an open question whether there exists a topological index to count 
the number of moduli directly, as has been done for moduli spaces of 
self-dual instantons, monopoles, and other BPS states in SUSY theories.

Last we note that the junctions described in this paper bear some
rough resemblance to string junctions of \cite{stringjunc}, and at first
one might guess that string junctions might be described, perhaps after
a series of duality transformations, by the field theory wall
junctions described here.
Even leaving aside the different amounts of supersymmetry
preserved by the two types of configurations -- eight supercharges versus
one -- the analysis of sections \ref{z3} and \ref{ldl} makes it clear
that the relation between the two, if any, cannot be too straightforward.
It is known that type $IIB$ string junctions 
can be arranged to form infinite network lattices, and a simple argument
shows that this cannot be the case for domain wall junctions in our theory
with $Z_3$-symmetric superpotential.  While one can arrange a hexagonally
symmetric configuration as described in \cite{gt}, such a configuration
cannot be BPS, since if one chooses to preserve a fixed supercharge,
the orientation of a domain wall segment is completely determined via the 
long-distance limit of the integral BPS equations in terms of the phase
of the superpotential difference across the segment, something which
does not hold for the hexagonal lattice.  The same considerations apply
to rectangular lattices in theories with the analogous $Z_4$-symmetric
superpotential.

One can imagine, of course, evading this no-go principle for BPS lattices
by choosing the superpotential to be of the form $W(\phi) = p(x) + k\cdot
\phi$, where $p(\phi)$ is some holomorphic, doubly periodic function in
the compex plane.  One could then construct formal long-distance limits
of junction lattices which preserve supersymmetry.  However since doubly
periodic holomorphic functions of one variable are necessarily singular,
one would have to resolve the singularities with new degrees of freedom
in order to give any physical meaning to such configurations, and we
do not pursue this problem here.

\section{Geometric reverse-engineering: an explicit construction}
\label{z4}

Despite the consistent picture we have presented of 1/4-BPS
domain wall junctions, a skeptic might still suspect that there could be
a subtle obstruction to the existence of such a state.  In order to
allay any such anxieties, we will demonstrate that it is straightforward
to construct $1/4$-supersymmetric configurations describing
wall junctions and networks thereof.

We begin with the observation that the statement $ \Omega^* \partial_z W 
=  \Omega \partial_{\bar{z}} W^*$ for some phase $\Omega$
always holds for $1/4$-BPS states, independent of the form of the
K\"ahler metric.
In other words, the condition for a function $W(z,\bar{z})$
describing the behavior of the superpotential as a function of space
is simply the condition that $(W^*, W)$ be equal to $(\Omega^* \partial_z A,
\Omega \partial_{\bar{z}} A)$ for some real function $A$.

Furthermore, if we know more or less what the energy density of
the state should look like, we can simply construct the function $A$
by solving the linear Poisson equation $\partial_z \partial_{\bar{z}}
A = {1\over 4} V(z,\bar{z})$.  Having solved for the superpotential
as a function of space, the value of the field $\phi$ is then implicitly
defined.

There are two catches to this procedure.  The first is that one cannot
start with a given K\"ahler metric and use this procedure to solve the
BPS equations for that metric.  Once one has a profile for the
field $\phi$ one can, of course, reconstruct the corresponding
metric, if not necessarily in closed form, although one must then
check that the resulting metric is nonsingular and positive definite.
We shall give an example below.
The second catch is that the process
of solving for $\phi(z,\bar{z})$ in terms of the superpotential
$W(z,\bar{z})$ can break down if the function $W(z,\bar{z})$ ever attains
a value for which $W^\prime(\phi)$ vanishes.
These caveats, however, are really blessings in disguise.  If not
for such obstructions, one could clearly start with energy density
distributions
with no reasonable interpretation as a junction state or a network thereof,
and use them to generate 1/4-BPS states.

There is another, more subtle, property that the function $W(z,\bar{z})$
must satisfy for a BPS junction or network.  The BPS equation
for a single static wall,
\be
K_{\phi \phi^*} \vec{n} \cdot\vec{\nabla} \phi = \omega W^{\prime}(\phi)^*,
\ee
where $\vec{n}$ is the direction normal to the wall and $\omega$ is a phase,
means that the imaginary part of $\omega W^*$ remains constant over a
BPS wall trajectory -- that is, the trajectory in the $W$-plane of a
single domain wall solution is always a straight line segment connecting
two vacua \cite{fmvw}.
This implies a restriction on $W(z,\bar{z})$ for a BPS junction
or network state:
that the set of values assumed by $W$ over the $z-\bar{z}$ plane must
exactly fill out the convex hull in the $W$-plane of the $k$ vacua among
which the junction interpolates.  That is, the image in the $W$-plane
of the junction with $k$ legs is simply a $k$-sided polygon, which we
will refer to henceforth as the \it BPS polygon \rm in the $W$-plane.

To make this less abstract, it would be nice to construct explicitly
the functions $W(z,\bar{z})$
and $A(z,\bar{z})$ for the basic three-wall junction
in the $Z_3$ theory.  However we have not been able to find an energy
density profile which allowed us to solve the Poisson equation in
closed form for such a configuration.
Instead, we will turn to a case in which a simple
closed-form solution does exist, and
verifiably has the correct properties to describe a well-behaved
junction state: the $Z_4$-symmetric four-wall junction in the $Z_4$ theory
which lies at the intersection of the $s$-channel and $t$-channel branches
of the moduli space of four-wall networks in that theory \cite{t&v}.

Given the superpotential
\be
  W (\Phi) = \Lambda^2 \Phi - {{\Phi^5}\over {5\Lambda^2}}, 
  \label{z4w}
\ee
which has vacua at $\pm \Lambda, \pm i \Lambda$, there is a natural guess
at a profile function $W(z,\bar{z})$ for a $Z_4$-symmetric four-wall junction:
\be
  W = {{(2-2i) \Lambda^3}\over 5}
  \left ( \tanh\{ \Lambda (z+\bar{z}) \} - i \tanh\{i \Lambda(z - \bar{z})\}
  \right)
\ee
Note that our initial profile does indeed map to itself under a 
combined discrete spatial rotation and $R$-symmetry transformation, as we
expect the actual solution to do.
We now verify that this function has the correct behavior to be a
junction state.

First consider the behavior of $W$ as $x\rightarrow +\infty$, and $y = mx$,
with $m > 0$.  In this limit 
\be
W\rightarrow {{(2-2i) \Lambda^3}\over 5} ( 1 + i) = {{4 \Lambda^3}\over 5} \ ,
\ee
which is the correct value of the superpotential in the vacuum at $\phi
= \Lambda$.
Since we know how $W$ transforms under the discrete $Z_4$ rotation subgroup, 
this then implies that $W$ has the correct limiting behavior in all four 
quadrants of the $z-\bar{z}$ plane.

\begin{figure}
  \centerline{\epsfbox{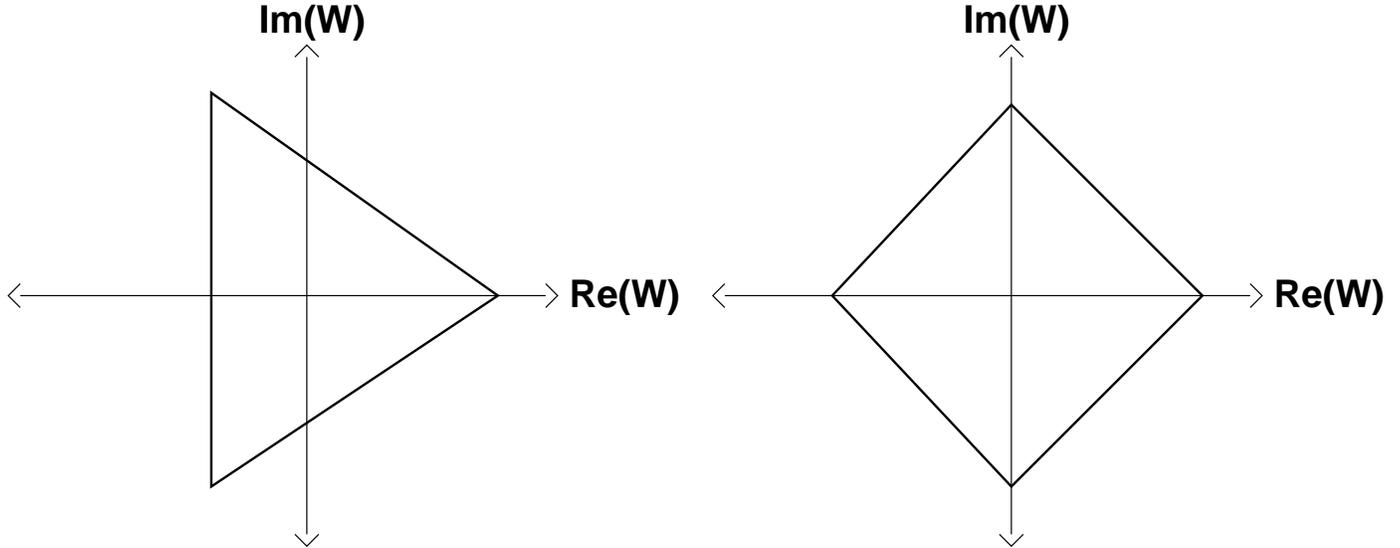}}
  \vspace{0.3in}
  \caption{\sf BPS polygons in the $W$-plane for the two examples discussed in
   this section.  Vertices (vacua) preserve
   four supercharges, edges (half-walls) preserve two, and the interior
   (the junction) preserves one.}
  \label{stwall2}
\end{figure}

Second, it is clear that the set of values assumed by $W$ is precisely the
convex hull in the $W$ plane of the four vacuum values of the
superpotential (see figure \ref{stwall2})  
so the solution's $W$-profile does indeed fill out the BPS polygon
whose edges are 
the trajectories of the superpotential along the four individual BPS domain 
(half-) wall solutions.
Since $W(z,\bar{z})$ stays within the BPS polygon for all $z,\bar{z}$,
the function $\phi(z,\bar{z})$ can be recovered from equation~(\ref{z4w}), 
since $W'(\phi)$ is non-vanishing everywhere in this region.

Until now we have let the K\"ahler potential be an arbitrary function
$K(\Phi, \Phi^\dagger)$. Then the condition for a junction state to
preserve the supercharge $Q_\uparrow - \Omega Q_\uparrow^\dagger$ is

\be
2 K_{\phi \phi^*} \partial_z \phi = \Omega\left( \Lambda^2 - 
\frac{\phi^{*4}}{\Lambda^2} \right)\ .
\ee
Therefore, the K\"ahler metric can be expressed as
\be
K_{\phi \phi^*} (z,\bar{z}) =  \Omega \left[\frac{\Lambda^2 - 
\phi^{*4} /\Lambda^2}{2 \partial_z \phi}\right] \ .
\label{kahler}
\ee
For this to be a sensible metric, the function $K_{\phi\phi^*}$
must be real and non-degenerate.  To check this, we point out that
the requirement that $K_{\phi \phi^*}$ be real is equivalent to
the requirement that
\be
\Omega^* (\partial_z \phi) W^\prime(\phi) = \Omega(\partial_{\bar{z}} \phi^*)
W^\prime (\phi)^*
\ee
However, the superpotential can be written as 
$W = \Omega \partial_{\bar{z}} A(z,\bar{z})$ where $\Omega =
\exp ( - {{\pi i}\over 4} )$ and $A$ is the real function

\be
A = {{2 \sqrt{2} \Lambda^2} \over 5} \left [
 \ln \cosh ( \Lambda (z + \bar{z}) ) + \ln \cosh ( i \Lambda  (z - \bar{z}))
\right ]
\ ,
\ee
and we may therefore infer that $K_{\phi \phi^*}(z,\bar{z})$
is indeed everywhere
real.  Moreover, as the gradient of $W$ is everywhere non-vanishing in the
interior of the BPS polygon, this means that $K_{\phi\phi^*}$
is real and nonzero, and thus positive definite
for all $z,\bar{z}$.  Therefore, since our theory contains only one chiral 
superfield, the metric defined in this way satisfies the K\"ahler condition 
trivially.

Hence, we have demonstrated that, for some choice of reasonable 
metric and the superpotential (\ref{z4w}),
there exists a domain wall junction preserving a single supercharge.
It remains to note that the map from the $(z,\bar{z})$ plane
to the BPS polygon is nonsingular and one-to-one. Thus
we could formally re-express the metric as a positive definite function of 
$\phi$ and $\phi^*$, which could be derived from a K\"ahler potential 
$K_{\rm initial}$.

Having developed this existence proof for nontrivial K\"ahler metric, we may
wonder again about the case with trivial metric. At present we have a partial 
answer to this question. Let
\be
K(t;\phi, \phi^*) \equiv (1-t) K_{initial} (\phi, \phi^*)
+ t\phi\phi^* \ .
\ee
Then the corresponding metric is positive definite $\forall \ t\in[0,1]$.
We expect the number of supersymmetric configurations in a given topological
class to remain the same as we vary the parameters of the Lagrangian in a
sufficiently smooth way.
And so as we vary the metric, we can vary the solution to the BPS
equations, 
knowing that a solution must exist for all values of $t$, including $t=1$.
Therefore it seems that a solution exists for trivial K\"ahler metric, and 
our argument is complete.

While it is still possible that there is some obstruction to continuing
the solution to $t=1$, other than an ill-defined metric, at very least, we 
have shown that a metric can be constructed which allows
BPS junctions.

\section{Relevance to $N=1$ SYM}
\label{sym}

The models we wrote down were quite simple, based on a single chiral
superfield and a polynomial superpotential.  However they are closely related
to a particular physical system of great interest.
The simplest four-dimensional
supersymmetric gauge theories, pure super-Yang-Mills theories with no
matter multiplets, are believed to have a finite number of equivalent
supersymmetric vacua ($n$ of them for the gauge group $SU(n)$ \cite{witcon})
related by a spontaneously broken discrete $R$-symmetry, as does our model.

The picture is as follows.  The $SU(n)$ theory has, at the classical
level, a $U(1)$ $R$-symmetry under which the gauge fields transform
trivially and the gluinos $\lambda_\alpha^a$ have charge $+1$.  The
Dynkin index of the adjoint representation is equal to $2n$, so the 
$R$-symmetry
is broken by anomalies down to a $Z_{2n}$ subgroup.  Since bosonic
operators transform under a $Z_n$ subgroup, strong coupling effects can
at most spontaneously break $Z_{2n}$ to $Z_2$, so 
the maximum number of supersymmetric vacua the theory may have is $n$.
An index calculation \cite{witcon} shows that this upper bound is attained.

The existence of disconnected vacua means that $SU(n)$ SYM has domain walls
interpolating between them, and the anomaly term (\ref{wallanom}) makes it
possible for the domain walls to preserve half the supersymmetry.  Indeed,
this is the context in which BPS-saturated domain walls were originally
discovered.  Can they form BPS junctions?  And if so, can we calculate their
tensions explicitly?

The description of the effective theory of SYM is quite complicated,
and at present still poorly understood. 
But the considerations of sections (\ref{ldl}) and (\ref{z4})
suggest that within the category of
SUSY theories with multiple vacua related by a
spontaneously broken cyclic $R$-symmetry, domain wall junction and network
states are a fairly robust feature whose existence depends
little on the details of the metric.  Indeed, the only essential input to
our calculation was that the theory contain a single chiral superfield.

The authors of \cite{dgk},\cite{dvalkak}, building on
earlier work \cite{vy}, derive an
effective action for large-$n$ SUSY gluodynamics in its confining phase.
Their model describes spontaneous $Z_n$ breaking and domain walls
in terms of a single superfield $X$, which is closely related
to the gluino condensate $\langle\lambda\lambda\rangle$.
The superpotential for this model,
\be
  W_{eff} = n \Lambda^2 X - {{CNX^{n+1}}\over {n+1}} \ ,
\ee 
(where $C$ is a constant), clearly gives the correct global behavior for the
system: the superpotential transforms nontrivially under a spontaneously
broken $Z_n$ $R$-symmetry, and there are BPS domain walls interpolating
between distinct vacua.
Indeed, this superpotential is trivially related to ours: by rescaling
$X$, one can make the two superpotentials the same.  The explicit dependence
of $W_{eff}$
on $C$ and $n$ is then absorbed into the normalization of the metric.

We therefore expect that:
\\
\\
{\it $SU(n)$
SYM contains in its Hilbert space wall junction states preserving
a single Hermitian supercharge, at least for sufficiently large $n$.}
\\

The tension of the walls was calculated in \cite{a&t,dvalshif}.  In order to
calculate the additional energy contributed by the junction itself,
one would need to use the K\"ahler metric for this model, which, unfortunately,
is not known.  So we can at present make a qualitative prediction about the
existence of $1/4$-supersymmetric
junctions in $N=1$ SYM, but not yet a quantitative one about their
tensions.

\section{Discussion}
\label{discussion}

SYM with gauge group $SU(n)$ is believed to have a dual formulation in
terms of $M$-theory.  In \cite{wittenqcd}, Witten describes a configuration
of M-theory fivebranes with a three-dimensional intersection, the effective
field theory on which is argued to be in the same universality class
as pure $N=1$ supersymmetric Yang-Mills theory.  Many of the strong-coupling
phenomena present in SYM, such as color flux tubes and domain walls, then
have simple descriptions in terms of intersecting branes in M-theory.
Indeed, explicit M-theory states have been written down
which correspond to BPS domain walls in this effective theory
\cite{fayspal}, \cite{volov}.

We expect that there ought to exist solutions of 11-dimensional
supergravity preserving $1/32$ of the supersymmetry of the theory,
corresponding to the full spectrum of BPS wall junctions described in this
paper.
The question of the existence of such solutions therefore may
provide a stringent and quantitative test of the duality between $M$-theory
and $N=1$ SYM.

\section*{Acknowledgments}

The authors would like to thank Oren Bergman, Zurab Kakushadze, 
Andrew Sornborger and Cyrus Taylor for
valuable discussions.  This work was supported in part by the National 
Science Foundation under grants PHY/94-07194 and PHY/97-22022, and by 
the U.S. Department of Energy (D.O.E.).


\begin{thebibliography}{999}
\parindent=.6em
\bibitem{a&t} E.R.C. Abraham and P.K. Townsend, {\it Nucl. Phys.} {\bf B351},
313 (1991).

\bibitem{dvalshif} G. Dvali and M. Shifman, {\it Phys. Lett.} {\bf B396},
64 (1997); {\tt hep-th/9612128}.

\bibitem{rps} B.S.~Ryden, W.H.~Press and D.N.~Spergel, {\it Astrophys.
Journ.} {\bf 357}, 293 (1990).

\bibitem{ct} S.M. Carroll and M. Trodden, {\it Phys. Rev. D} {\bf 57},
5189 (1998); {\tt hep-th/9711099}.

\bibitem{t&v} S.V. Troitsky and  M.B. Voloshin, {\it Phys. Lett.} {\bf B449},
17 (1999).

\bibitem{gt} G.W. Gibbons and P.K. Townsend, {\tt hep-th/9905196}.

\bibitem{witcon} E. Witten, {\it Nucl. Phys.} {\bf B202}, 253 (1982).

\bibitem{extend} J. A. de Azcarraga, J. P. Gauntlett, J. M. Izquierdo and
P. K. Townsend, {\it Phys. Rev. Lett.} {\bf 63}, 2443 (1989).

\bibitem{cosmic} S.C. Davis, A.-C. Davis, and M. Trodden, {\it Phys. Lett.}
{\bf B405}, 257 (1997); {\tt hep-ph/9702360}.

\bibitem{stringjunc} J. H. Schwarz, {\it Nucl. Phys. Proc. Suppl.}
{\bf B55}, 1 (1997); {\tt hep-th/9607201}.

\bibitem{fmvw} P. Fendley, S.D. Mathur, C. Vafa and N.P. Warner, {\it Phys.
Lett.} {\bf B243}, 257 (1990).

\bibitem{dgk} G. Dvali, G. Gabadadze, and Z. Kakushadze, {\tt hep-th/9901032}. 

\bibitem{dvalkak} G. Dvali and Z. Kakushadze, {\it Nucl. Phys.}
{\bf B537}, 297 (1999); {\tt hep-th/9807140}.

\bibitem{vy} G. Veneziano and S. Yankielowicz, {\it Phys. Lett.} 
{\bf 113B}, 231 (1982).

\bibitem{wittenqcd} E. Witten, {\it Nucl. Phys.} {\bf B507}, 658 (1997);
{\tt hep-th/9706109}.

\bibitem{fayspal} A. Fayyazuddin and M. Spalinski, {\tt hep-th/9711083}.

\bibitem{volov} A. Volovich, {\tt hep-th/9710120}; A. Volovich, {\it Phys.
Rev.} {\bf D59}, 065005 (1999); {\tt hep-th/9801166}.

\end{thebibliography}
\end{document}